\documentclass[conference]{IEEEtran}
\IEEEoverridecommandlockouts
\usepackage{cite}
\usepackage{amsmath,amssymb,amsfonts}
\usepackage{algorithmic}
\usepackage{textcomp}
\usepackage{xcolor}
\usepackage[pdftex]{graphicx}
\usepackage{epstopdf}
\usepackage{color}

\usepackage{amsmath,amssymb,enumerate} 
\usepackage{bm} 
\usepackage{indentfirst} 
\usepackage{empheq} 
\usepackage{here} 
\newcommand{\dd}{\mathrm{d}}

\newcommand{\ii}{\mathrm{i}}

\def\BibTeX{{\rm B\kern-.05em{\sc i\kern-.025em b}\kern-.08em
    T\kern-.1667em\lower.7ex\hbox{E}\kern-.125emX}}
\begin{document}

\title{Independence of the Fundamental Equation of the Oscillation Model on Algebraic Representations: Social Media Echo Chamber Effect}

\author{\IEEEauthorblockN{Kakeru  Ohki}
    \IEEEauthorblockA{
        \textit{Tokyo Metropolitan University} \\
Tokyo 191--0065, Japan \\
ohki-kakeru@ed.tmu.ac.jp}
\and
\IEEEauthorblockN{Ayako Hashizume}
    \IEEEauthorblockA{
        \textit{Hosei University} \\
Tokyo 194--0298, Japan \\
hashiaya@hosei.ac.jp}
\and
\IEEEauthorblockN{Masaki Aida}
    \IEEEauthorblockA{
        \textit{Tokyo Metropolitan University} \\
Tokyo 191--0065, Japan \\
aida@tmu.ac.jp}
}

\maketitle

\begin{abstract}
In the oscillation model that describes the user dynamics of online social networks, it is known that the fundamental equation can explicitly describe the causal relationship between the network structure and user dynamics.
The fundamental equation uses algebra that satisfies the anti-commutation relation, and its matrix representation is not unique.
However, even if the matrix representations are different, the same results should be derived from different representations of the fundamental equation if they are describing the same phenomenon.
In this paper, we confirm, using the echo-chamber effect as an example, that the fundamental equations of different matrix representations lead to the same result.
\end{abstract}


\section{Introduction}
With the rapid development of information networks in recent years, social networking services (SNSs) have become widespread, and user interest in online social networks (OSNs) has become active.
While such user dynamics in OSN support human activities in the real world by promoting information exchange and mutual understanding, they also cause social problems such as the online flaming phenomenon and echo-chamber effect.
Therefore, understanding user dynamics in OSN is a crucial issue.

The oscillation model is known to be able to describe user dynamics on OSNs~\cite {Aida2018,Aida2016}.
This is a minimal model that applies the wave equation on networks.
The wave equation is an equation that expresses the phenomenon that an object propagates through a medium at finite speed.
In the concept of the oscillation model, the strength of the users' activity should have some influence on each other through the OSN, so the wave equation models the propagation of the influence through the OSN at finite speed.

The concept of node centrality represents the strength and/or importance of the activity of nodes in the network \cite{wasserman,carrington,mislove}.
The characteristic of the oscillation model is that the oscillation energy gives a generalized concept of node centrality and so can express the strength of network activity \cite{Takano2018,Takano2016}.
Conventionally, typical node centralities used in network analysis are the degree centrality and the betweenness centrality; fortunately, the oscillation model can provide a unified framework for both of them.
That is, under the condition that there is no bias in network usage, all link weights are assumed to be $ 1 $, the oscillation energy of each node yields degree centrality. 
On the other hand, taking the link weight to be the number of shortest paths that pass through that link (e.g., the volume of traffic), the oscillation energy of each node leads to the betweenness centrality.
Furthermore, the oscillation model can be applied even when the network usage is biased, and provides a generalization of node centrality.

The oscillation model makes it possible to derive a fundamental equation that can explicitly describe the causal relationship between the structure of the network and the user dynamics generated from the structure.
However, since the method of representing the fundamental equations using a matrix is not unique, it is necessary to confirm whether the fundamental equations in different matrix representations describing the same phenomenon yield the same results.
If we get the same result, the fundamental equation can be considered to be {\em well-defined}. 
This paper takes the echo-chamber effect as an example and shows that the same user dynamics can be derived from the fundamental equations in different matrix representations.

The rest of this paper is organized as follows.
Sec.~\ref{sec:OM} gives an overview of the oscillation model and explains the fundamental equation.
In Sec.~\ref{sec:E-C}, we show the characteristics of user dynamics representing the echo-chamber effect that is derived by using the fundamental equation.
In Sec.~\ref{sec:main}, another representation of the fundamental equation is introduced, the user dynamics representing the echo-chamber effect is derived by using the other representation. 
The results show that the same user dynamics as described in Sec.~\ref{sec:E-C} are obtained.
Finally, Sec.~\ref{sec:conclusion} concludes this paper.

\section{Oscillation Model for User Dynamics in OSNs}
\label{sec:OM}
The fundamental equations of user dynamics can be summarized as follows.
First, for nodes $i,\,j \in V$ of directed graph $G (V, E)$ that represents the structure of an OSN with $n$ users, if the weight of directed link $(i \rightarrow j) \in E$ is given as $ w_ {ij} $, the adjacent matrix $\bm{\mathcal{A}} = [\mathcal{A}_{ij}]_{1\le i,j \le n}$ is defined as 
\begin{align}
\mathcal{A}_{ij} := \left\{
\begin{array}{cl}
w_{ij},&  \quad (i\rightarrow j) \in E,\\
0,& \quad (i\rightarrow j) \not\in E.
\end{array}
\right. 
\end{align}
Also, given nodal (weighted) out-degree $d_i := \sum_{j\in \partial i} w_{ij}$, 
the degree matrix is defined as 
\begin{align}
\bm{\mathcal{D}} := \mathrm{diag}(d_1,\,\dots\,d_n).
\end{align}
Here, $\partial i$ denotes the set of adjacent nodes of out-links from node $i$. 
Next, the Laplacian matrix of the directed graph representing the structure of the OSN is defined 
by 
\begin{align}
\bm{\mathcal{L}} := \bm{\mathcal{D}} - \bm{\mathcal{A}}.
\end{align}

Let the state vector of users at time $t$ be 
\[
\bm{x}(t) := {}^t\!(x_1(t),\,\dots,\,x_n(t)), 
\]
where 
$x_i(t)$ $(i=1,\,\dots,\,n)$ denotes the user state of node $i$ at time $t$. 
Then, the wave equation for the OSN is written as 
\begin{align}
\frac{\dd^2}{\dd t^2} \, \bm{x}(t) = - \bm{\mathcal{L}} \, \bm{x}(t). 
\label{eq:wave-eq}
\end{align}
Here, in addition to simply finding the solution $\bm{x}(t)$ of the wave equation (\ref{eq:wave-eq}), 
it is desirable to be able to describe what kind of OSN structure impacts user dynamics. 
In other words, we want to describe the causal relationship between OSN structure and user dynamics. 
To achieve this, we need to develop a first-order differential equation with respect to time 
(hereinafter referred to as the fundamental equation)~\cite{Aida-book,Aida2020}.

Let us introduce a new matrix, $\bm{\mathcal{H}}$, as follows. 
\begin{align}
\bm{\mathcal{H}} := \sqrt{\bm{\mathcal{D}}^{-1}} \, \bm{\mathcal{L}} = \sqrt{\bm{\mathcal{D}}} - \sqrt{\bm{\mathcal{D}}^{-1}} \, \bm{\mathcal{A}}, 
\label{eq:def_H}
\end{align}
where $\sqrt{\bm{\mathcal{D}}} := \text{diag}(\sqrt{d_1},\,\dots,\,\sqrt{d_n})$. 
As is well-known, the normalized Laplacian matrix is defined as
\[
\bm{\mathcal{N}} := \sqrt{\bm{\mathcal{D}}^{-1}}\, \bm{\mathcal{L}} \,  \sqrt{\bm{\mathcal{D}}^{-1}} = \bm{I} -   \sqrt{\bm{\mathcal{D}}^{-1}} \, \bm{\mathcal{A}}\, \sqrt{\bm{\mathcal{D}}^{-1}}; 
\]
so we call $\bm{\mathcal{H}}$ the semi-normalized Laplacian matrix. 
Here, $\bm{I}$ is the $n\times n$ unit matrix. 

Using the semi-normalized Laplacian matrix $\bm{\mathcal{H}}$, a new fundamental equation for user dynamics can be written as follows.
\begin{align}
  \ii \, \frac{\dd}{\dd t} \,  \bm{\hat{x}}(t)
  = \left(\bm{\mathcal{H}} \otimes \bm{\hat{a}}
 + \bm{\sqrt{\mathcal{D}}} \otimes \bm{\hat{b}}
 \right)\bm{\hat{x}}(t), 
 \label{eq:fundamental1}
\end{align}
where $\bm{\hat{a}}$ and $\bm{\hat{b}}$ are $2\times 2$ matrices defined as 
\[
\bm{\hat{a}} = 
\frac{1}{2}
\begin{bmatrix}
 +1 & +1\\ 
 -1 & -1 
\end{bmatrix}, \quad 
\bm{\hat{b}} = 
\frac{1}{2}  
\begin{bmatrix}
 +1 & -1\\ 
 +1 & -1 
 \end{bmatrix} ,
\]
and $\bm{\hat{x}}(t)$ is the $2n$-dimensional state vector. 
The solution $\bm{x}(t)$ of the original wave equation (\ref{eq:wave-eq}) can be obtained from the solution $\bm{\hat{x}}(t)$ by 
\begin{align}
\bm{x}(t) = (\bm{I}\otimes (1,1))\,\bm{\hat{x}(t)}. 
\label{eq:x}
\end{align}
$\otimes$ denotes the Kronecker product~\cite{brewer}. 

The fundamental equation is similar to the Dirac equation of relativistic quantum mechanics, and its feature is that $\bm{\hat{a}}$ and $\bm{\hat{b}}$ satisfy the anti-commutation relation: 
\begin{align}
\{\bm{\hat{a}},\bm{\hat{b}}\} := \bm{\hat{a}}\,\bm{\hat{b}} + \bm{\hat{b}}\,\bm{\hat{a}} = \bm{\hat{e}}, \quad 
\bm{\hat{a}}^2 = \bm{\hat{b}}^2 = \bm{\hat{\mathrm{o}}},
\label{eq:anti-comm}
\end{align}
where $\bm{\hat{e}}$ denotes the $2\times 2$ unit matrix and $\bm{\hat{\mathrm{o}}}$ denotes the null matrix. 

\section{Echo-Chamber Effect}
\label{sec:E-C}
The echo-chamber effect is a phenomenon in which beliefs that are far from common sense are strengthened within a relatively small community existing in an online social network.
One model proposed to explain the occurrence of the echo-chamber effect posits that a relatively small partial network corresponding to a closed community detaches itself from the online social network, and the separated subnetwork becomes a complete graph \cite{ITC32,CQ202006}.

Here, based on the fundamental equation (\ref{eq:fundamental1}), let us consider the situation in which the weights of all the links in the separated subnetwork have the same value of $w$ \cite{ICETC2020}.
This is a situation in which the weight of the link, which indicates the strength of the relationship between users, has increased to the limit and is saturated because the discussion is activated in the divided community.
At this time, if the number of users in the community is $n$, the nodal degree is the same for all nodes, and $d = (n-1)\,w$.
Also, in the corresponding Laplacian matrix, all eigenvalues other than $0$ are duplicated, and the eigenvalues are denoted as $\lambda = \omega^2 = n\,w$.

In this situation, since the degree matrix is $\bm{\mathcal{D}} = d\,\bm{I}$, which is proportional to the identity matrix $\bm{I}$, the matrices $\bm{\mathcal{H}}$ and $\sqrt{\bm{\mathcal{D}}}$ appearing in the fundamental equation (\ref{eq:fundamental1}) can exhibit simultaneous diagonalization. 
Therefore, if the fundamental equation (\ref{eq:fundamental1}) is transformed so that the matrices $\bm{\mathcal{H}}$ and $\sqrt{\bm{\mathcal{D}}}$ are diagonalized simultaneously and expressed in the block diagonalized form of $2\times 2$, the equations for all the eigenvalues, other than the eigenvalue of $0$ of the Laplacian matrix, are the same and can be written as follows:
\begin{align}
\ii \, \frac{\dd}{\dd t} \bm{\psi}(t) 
&= \left(\frac{\omega^2}{2\,\sqrt{d}} 
  \begin{bmatrix}
   +1 & +1\\ 
   -1 & -1
  \end{bmatrix}
  + \frac{\sqrt{d}}{2} 
  \begin{bmatrix}
   +1 & -1\\ 
   +1 & -1
  \end{bmatrix}
  \right) \bm{\psi}(t)
\notag\\
&=
  \begin{bmatrix}
   +\frac{\omega^2+d}{2\,\sqrt{d}} & +\frac{\omega^2-d}{2\,\sqrt{d}}\\ 
   -\frac{\omega^2-d}{2\,\sqrt{d}} & -\frac{\omega^2+d}{2\,\sqrt{d}}
  \end{bmatrix}
\, \bm{\psi}(t).  
\label{eq:block-diag1}
\end{align}
Here, if the two-dimensional vector of the solution of (\ref{eq:block-diag1}) is denoted as  
\begin{align}
\bm \psi (t) = 
 \begin{pmatrix}
  \psi^+ (t) \\ 
  \psi^- (t)
 \end{pmatrix},
\end{align}
and set the Ansatz of (\ref{eq:block-diag1}) as 
\begin{align}
  \psi^\pm(t) := \exp\!\left(\mp \ii\theta^\pm (t)\right),
  \label{eq:ansatz}
\end{align}
in double sign correspondence; this yields
\begin{align}
\frac{\dd}{\dd t} \theta^\pm (t) = \frac{\omega^2 + d}{2 \sqrt{d}} + \frac{\omega^2 - d}{2 \sqrt{d}}\exp\!\left(\pm \mathrm{i}\left(\theta^+(t) + \theta^-(t)\right)\right). 
\label{eq:d_theta/dt_1}
\end{align}
Since $\theta^\pm(t)$ is a complex number in general, by substituting 
\[
\theta^\pm(t) = \mathrm{Re}[\theta^\pm(t)] + \ii \, \mathrm{Im}[\theta^\pm(t)]
\]
into (\ref{eq:d_theta/dt_1}), we obtain the temporal evolutions of the real and the imaginary parts of $\theta^\pm(t)$ as 
\begin{align}
&\frac{\dd}{\dd t} \, \mathrm{Re}[\theta^\pm(t)] 
\notag\\
&= \frac{\omega^2+d}{2\sqrt{d}} + C^\pm(t) \, \cos\!\left(\mathrm{Re}[\theta^+ (t)] + \mathrm{Re}[\theta^- (t)]\right) 
\notag\\
&= \frac{\omega^2+d}{2\sqrt{d}} 
\notag\\
&\qquad {}+ C^\pm(t) \, \sin\!\left(-\left(\mathrm{Re}[\theta^\mp (t)]-\frac{\pi}{2}\right) - \mathrm{Re}[\theta^\pm (t)]\right), 
\label{eq:real1}\\
&\frac{\dd}{\dd t} \, \mathrm{Im}[\theta^\pm(t)] 
= \pm C^\pm(t) \, \sin\!\left(\mathrm{Re}[\theta^+ (t)] + \mathrm{Re}[\theta^- (t)]\right),  
\label{eq:imaginary1}
\end{align}
where 
\begin{align}
C^\pm(t) := \frac{\omega^2 - d}{2 \sqrt{d}}\,\exp\!\left(\mp \left(\mathrm{Im}[\theta^+(t)] + \mathrm{Im}[\theta^-(t)]\right)\right). 
\label{eq:C}
\end{align}

From this result, the following properties of the solution can be predicted.
Note that the temporal evolution (\ref{eq:real1}) of the real part of $\theta^\pm (t)$ has a structure similar to that of the Kuramoto model, since $C^\pm(t) > 0$.
The difference from the Kuramoto model is that $C^\pm(t)$ is not a constant. 
Now, if $C^\pm(t)$ is large enough and phase synchronization occurs as in a Kuramoto model, the following states are realized: 
\[
\mathrm{Re}[\theta^+ (t)] + \mathrm{Re}[\theta^- (t)] = +\frac{\pi}{2}. 
\]
Along with this, the temporal change (\ref{eq:imaginary1}) of the imaginary part of $\theta^\pm (t)$ becomes
\begin{align}
\frac{\dd}{\dd t} \, \mathrm{Im}[\theta^\pm(t)] 
&= \pm C^\pm(t), 
\label{eq:result1}
\end{align}
because the $\sin$ function part of (\ref{eq:imaginary1}) becomes $+1$.
Therefore, $\theta^+(t)$ increases and $\theta^-(t)$ decreases with time.
In both cases, these changes increase the amplitude of $\psi^\pm$ according to the Ansatz (\ref{eq:ansatz}).
This result means that the activity of the user in the community is activated, and it is considered that this describes the occurrence of the echo-chamber effect.

\section{Verification of Another Representation of the Fundamental Equation}
\label{sec:main}
The fundamental equation (\ref{eq:fundamental1}) is one representation of the fundamental equation; another representation that satisfies the anti-commutation relation (\ref{eq:anti-comm}) is given by 
\begin{align}
  \ii \, \frac{\dd}{\dd t} \,  \bm{\hat{x}}(t)
  = \left(\bm{\mathcal{H}} \otimes \bm{\hat{b}}
 + \bm{\sqrt{\mathcal{D}}} \otimes \bm{\hat{a}}
 \right)\bm{\hat{x}}(t), 
 \label{eq:fundamental2}
\end{align}
in which the $2\times 2$ matrices $\bm{\hat{a}}$ and $\bm{\hat{b}}$ are exchanged.
The same as (\ref{eq:fundamental1}), the solution of (\ref{eq:fundamental2}) is also linked to the solution of the wave equation (\ref{eq:wave-eq}) by (\ref{eq:x}).
Therefore, there should be no problem in adopting either (\ref{eq:fundamental1}) or (\ref{eq:fundamental2}) as the fundamental equation.

In this section, we would like to confirm whether the echo-chamber effect derived from the fundamental equation (\ref{eq:fundamental1}) in the previous section has the same result as the fundamental equation (\ref{eq:fundamental2}) in the other representation.

As in the previous section, we consider the situation that the degree matrix, $\bm{\mathcal{D}} = d\,\bm{I}$, is proportional to the identity matrix. 
Therefore, matrices $\bm{\mathcal{H}}$ and $\sqrt{\bm{\mathcal{D}}}$ can be simultaneously diagonalized. 
If the fundamental equation (\ref{eq:fundamental2}) is expressed in the block diagonalized form of $2\times 2$, the equations for all eigenvalues, other than the eigenvalue of $0$ of the Laplacian matrix, are the same, and can be written as follows:
\begin{align}
\ii \, \frac{\dd}{\dd t} \bm{\psi}(t) 
&= \left(\frac{\omega^2}{2\,\sqrt{d}} 
  \begin{bmatrix}
   +1 & \textcolor{red}{-1}\\ 
   \textcolor{red}{+1} & -1
  \end{bmatrix}
  + \frac{\sqrt{d}}{2} 
  \begin{bmatrix}
   +1 & \textcolor{red}{+1}\\ 
   \textcolor{red}{-1} & -1
  \end{bmatrix}
  \right) \bm{\psi}(t)
\notag\\
&=
  \begin{bmatrix}
   +\frac{\omega^2+d}{2\,\sqrt{d}} & 
   \textcolor{red}{-}\frac{\omega^2-d}{2\,\sqrt{d}}\\ 
   \textcolor{red}{+}\frac{\omega^2-d}{2\,\sqrt{d}} & -\frac{\omega^2+d}{2\,\sqrt{d}}
  \end{bmatrix}
\, \bm{\psi}(t), 
\label{eq:block-diag2}
\end{align}
Here, in order to emphasize the difference from the previous section, the different parts are shown in \textcolor{red}{red}.
As in the previous section, the two-dimensional vector of the solution of (\ref{eq:block-diag2}) is denoted as
\begin{align}
\bm \psi (t) = 
 \begin{pmatrix}
  \psi^+ (t) \\ 
  \psi^- (t) 
 \end{pmatrix},
\end{align}
and set the Ansatz of (\ref{eq:block-diag2}) as 
\begin{align}
  \bm{\psi^\pm}(t) := \exp\!\left(\mp \ii\theta^\pm (t)\right),
  \label{eq:ansatz_2}
\end{align}
in double sign correspondence; this yields 
\begin{align}
\frac{\dd}{\dd t} \theta^\pm (t) = \frac{\omega^2 + d}{2 \sqrt{d}} \,\textcolor{red}{-}\, \frac{\omega^2 - d}{2 \sqrt{d}}\exp\!\left(\pm \mathrm{i}\left(\theta^+(t) + \theta^-(t)\right)\right). 
\label{eq:d_theta/dt_2}
\end{align}
Since $\theta^\pm(t)$ is a complex number in general, by substituting \[
\theta^\pm(t) = \mathrm{Re}[\theta^\pm(t)] + \ii \, \mathrm{Im}[\theta^\pm(t)]
\]
into (\ref{eq:d_theta/dt_2}), we obtain the temporal evolutions of the real and the imaginary parts of $\theta^\pm(t)$ as 
\begin{align}
&\frac{\dd}{\dd t} \, \mathrm{Re}[\theta^\pm(t)] 
\notag\\
&= \frac{\omega^2+d}{2\sqrt{d}} \,\textcolor{red}{-}\, C^\pm(t) \, \cos\!\left(\mathrm{Re}[\theta^+ (t)] + \mathrm{Re}[\theta^- (t)]\right) 
\notag\\
&= \frac{\omega^2+d}{2\sqrt{d}} 
\notag\\
&\qquad {}+ C^\pm(t) \, \sin\!\left(-\left(\mathrm{Re}[\theta^\mp (t)]\,\textcolor{red}{+}\,\frac{\pi}{2}\right) - \mathrm{Re}[\theta^\pm (t)]\right), 
\label{eq:real2}\\
&\frac{\dd}{\dd t} \, \mathrm{Im}[\theta^\pm(t)] 
= \textcolor{red}{\mp} C^\pm(t) \, \sin\!\left(\mathrm{Re}[\theta^+ (t)] + \mathrm{Re}[\theta^- (t)]\right), 
\label{eq:imaginary2}
\end{align}
where $C^\pm(t)$ is defined as (\ref{eq:C}), the same as in the previous section.  

From this result, the following properties of the solution can be predicted.
Note that the temporal evolution (\ref{eq:real1}) of the real part of $\theta^\pm (t)$ also has a structure similar to that of the Kuramoto model.
Now, if phase synchronization occurs as in a Kuramoto model, the following states are realized: 
\[
\mathrm{Re}[\theta^+ (t)] + \mathrm{Re}[\theta^- (t)] = \textcolor{red}{-\frac{\pi}{2}}
\]
This is different from the previous section. 
Along with this, the temporal change (\ref{eq:imaginary2}) of the imaginary part of $\theta^\pm (t)$ becomes 
\begin{align}
\frac{\dd}{\dd t} \, \mathrm{Im}[\theta^\pm(t)] 
&= \textcolor{red}{\mp} C^\pm(t) \, \sin\!\left(\mathrm{Re}[\theta^+ (t)] + \mathrm{Re}[\theta^- (t)]\right) 
\notag\\
&= \textcolor{red}{\mp} C^\pm(t) \times (\textcolor{red}{-1})
\notag\\
&= \pm C^\pm(t),
\end{align}
because the $\sin$ function part of (\ref{eq:imaginary2}) becomes \textcolor{red}{$-1$}. 
This result is the same as (\ref{eq:result1}), and it can be seen that the behavior of the solution does not depend on the representation of the fundamental equation.

\section{Conclusion}
\label{sec:conclusion}
In the oscillation model, the fundamental equation can describe user dynamics, and at the same time, describe the causal relationship between the network structure and the user dynamics.
Since the fundamental equation is based on an algebraic structure that satisfies the anti-commutation relation and its matrix representation is not unique, there are two different representations in the fundamental equation.
The question is, do the different representations describe the same user dynamics for the same phenomenon.
In other words, if different results are derived from the fundamental equations, the properties of the solutions that appear will differ depending on the representations of the fundamental equation, which means that the theoretical model will contain contradictions.
In this paper, we compared the user dynamics derived from both fundamental equations for the echo-chamber effect and confirmed that they yielded the same user dynamics.

\section*{Acknowledgement}
This research was supported by Grant-in-Aid for Scientific Research 19H04096 and 20H04179 from the Japan Society for the Promotion of Science (JSPS).

\newpage


\end{document}